\documentstyle[11pt,newpasp,twoside,epsf]{article}
\def\edcomment#1{\iffalse\marginpar{\raggedright\sl#1\/}\else\relax\fi}
\reversemarginpar

\begin{document}
\title{High Resolution Spectra of Quasar AALs: \ 3C 191}
 \author{Fred Hamann}
\affil{Department of Astronomy, University of Florida, 
211 Bryant Space Science Center, Gainesville, FL 32611-2055}
\author{T.A. Barlow}
\affil{Infrared Processing and Analysis Center,
California Institute of Technology, 
MS 100-22, 770 South Wilson Ave., Pasadena, CA 91125}
\author{F.C. Chaffee}
\affil{California Association for Research in Astronomy, W.H. Keck
Observatory, 65-1120 Mamalahoa Highway, Kamuela, HI 96734}
\author{C.B. Foltz}
\affil{MMT Observatory, University of Arizona, 933 North Cherry Ave., 
Tucson, AZ 85721-0065}
\author{R.J. Weymann}
\affil{Observatories of the Carnegie Institution of Washington, 
813 Santa Barbara Street, Pasadena, CA 91101-1292}

\begin{abstract}
We discuss new high-resolution (6.7 km/s) spectra 
of the associated absorption lines (AALs) in the radio-loud quasar 
3C 191. The measured AALs have ionizations ranging 
from Mg I to N V, and  
multi-component profiles that are blueshifted by $\sim$400 
to $\sim$1400~km/s relative to the quasar's broad emission lines. 
Excited-state 
absorption lines of Si II$^*$ and C II$^*$ imply volume densities 
of $\sim$300 cm$^{-3}$ and a nominal distance from the quasar of 
28 kpc (assuming photoionization). The total column density is 
$N_H\sim 2\times 10^{20}$ cm$^{-2}$. Surprisingly, the absorber only 
partially covers the quasar emission source along our line of sight. 
We propose a model for the absorber in which pockets of dense neutral 
gas are surrounded by bigger clouds of generally lower density and 
higher ionization. This outflowing material might be leftover from a 
blowout associated with a nuclear starburst, the onset of quasar 
activity, or a past broad absorption line (BAL) wind phase.
\end{abstract}

\section{Introduction}

Associated absorption lines (AALs) are important diagnostics of the 
gaseous environments of quasars and active galactic nuclei (AGNs). 
In particular, the lines can trace a variety of phenomena 
--- from energetic outflows like the BALs to relatively 
quiescent gas at large galactic or inter-galactic distances 
(Weymann et al. 1979, Hamann \& Brandt 2001). 
We are involved in a multi-wavelength program to locate 
individual AAL absorbers, determine their elemental abundances, 
quantify their kinematic and physical 
properties, and understand the role of the AGNs and/or host 
galaxies in providing the source of material and kinetic 
energy to the absorbing gas. 

One interesting property is that, among radio-loud quasars, 
AALs appear more frequently and with greater strength 
in sources with ``steep'' radio spectra and/or 
lobe-dominated radio morphologies (Wills et al. 1995, Richards 2001, 
Brotherton et al. 1998). 3C 191
(Q0802+103, emission-line redshift $z_e = 1.956$) is a radio-loud 
quasar having both strong AALs and a bipolar, lobe-dominated radio 
structure. 
It provides a rare opportunity to define the 
distance between the quasar and 
the absorbing gas because its AALs include excited-state lines,  
e.g. C II$^*$ $\lambda$ 1336 and 
Si II$^*$ $\lambda\lambda$ 1265,1533, which constrain 
the volume density and therefore the quasar--absorber 
distance (assuming photoionization, Bahcall et al. 1967). 

\begin{figure}
\plotone{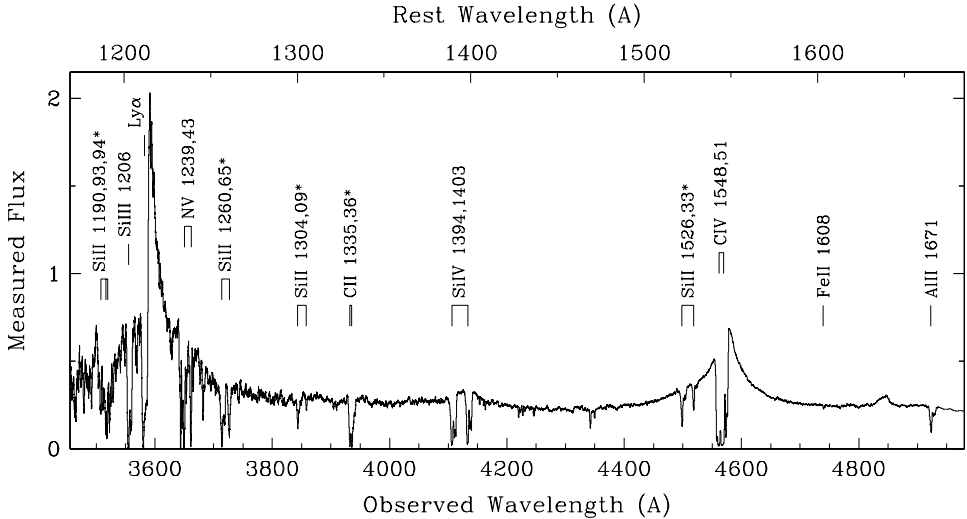}
\caption{Part of the 3C 191 spectrum showing its 
  strong associated absorption lines (labeled above). 
  The flux has units 10$^{-15}$ ergs cm$^{-2}$ s$^{-1}$ \AA$^{-1}$}
\end{figure}

\begin{figure}
\plotone{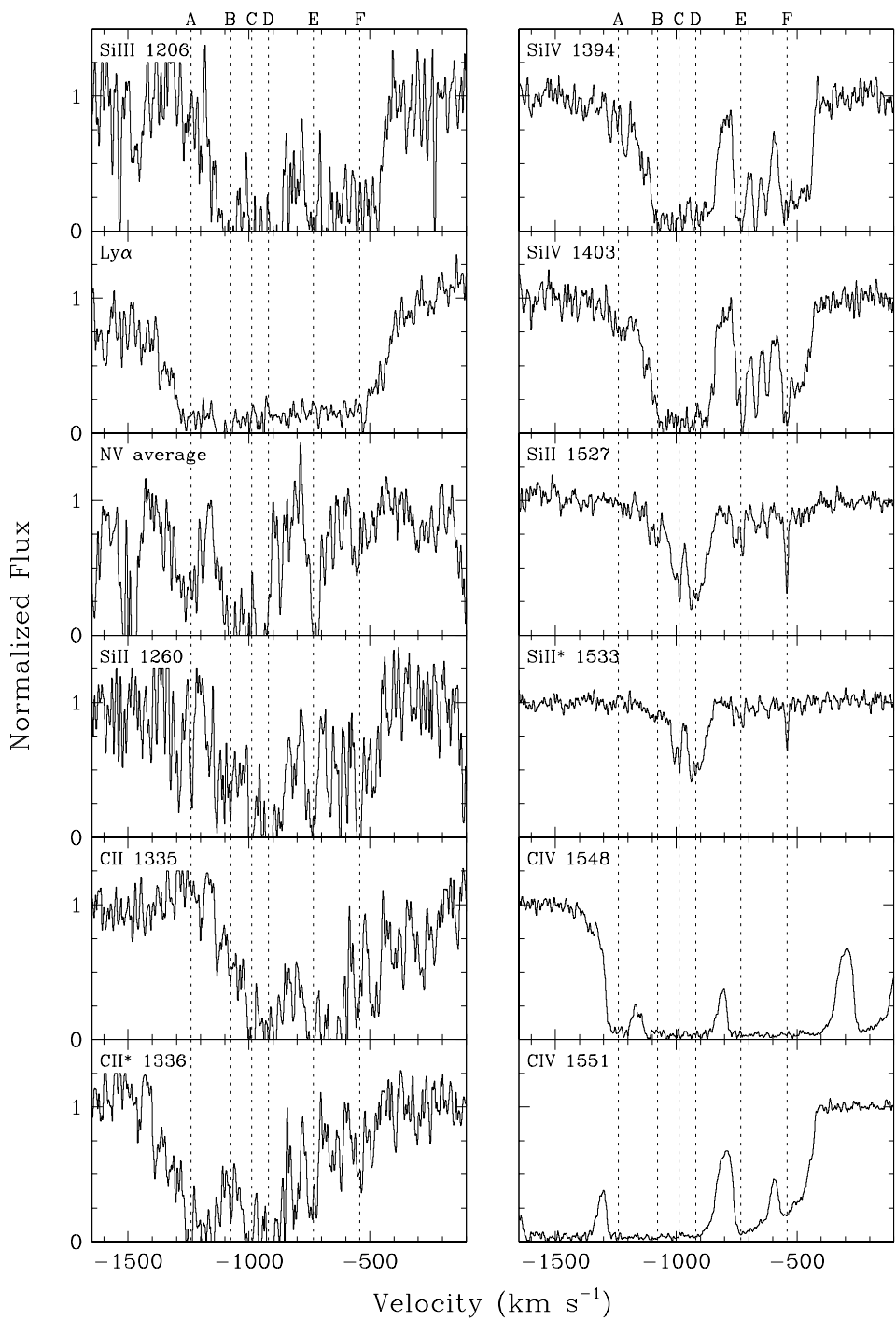}
\caption{Velocity profiles of the AALs in 3C 191.  
  Zero velocity corresponds to the nominal emission-line redshift, 
  $z_e = 1.956$. Dotted vertical lines mark the positions of strong 
  features.}
\end{figure}

\begin{figure}
\plotone{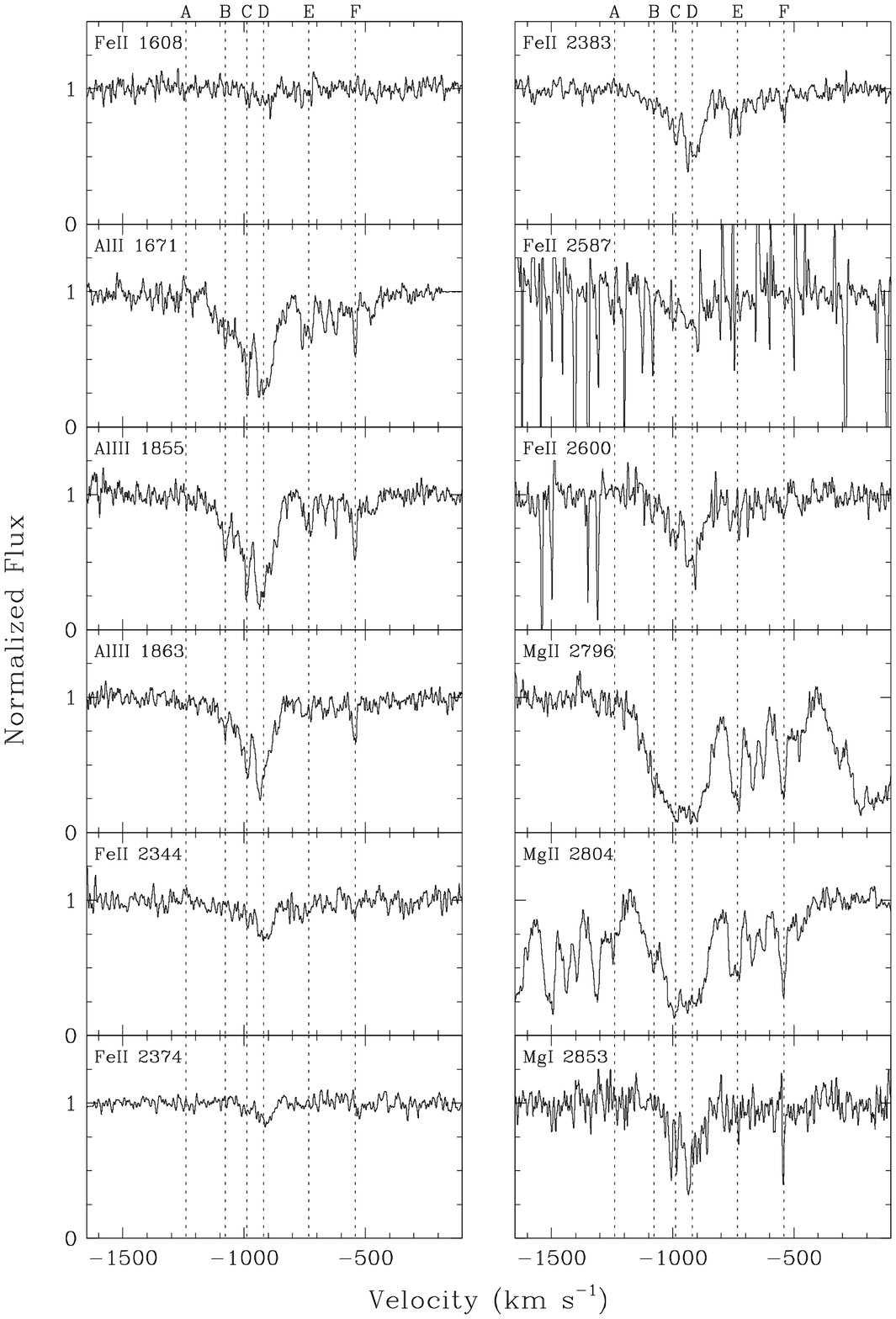}
\caption{See Figure 2.}
\end{figure}

\section{Observations and Results}

We observed 3C 191 on three occasions between 1997 and 1998 using the 
High Resolution Echelle Spectrograph (HIRES) on the Keck I telescope on 
Mauna Kea, Hawaii. On each occasion, a 0.86$^{\prime\prime}$ 
slit provided spectral
resolution $\lambda /\Delta\lambda \approx 45,000$ or 6.7 km s$^{-1}$, 
corresponding to 3 pixels on the 2048$^2$ Tektronix CCD. We reduced the data 
with standard techniques using the MAKEE software package. 
The spectra cover the 
observed wavelengths 3850 -- 5975 \AA\ and 6474 -- 8927 \AA. 

The measured AALs have ionizations ranging 
from Mg I to N V, and  
multi-component profiles that are blueshifted by $\sim$400 
to $\sim$1400~km/s relative to the quasar's broad emission lines 
(see Figures 1--3). These data yield the following new results.
\ {\bf (1)} The strengths of excited-state Si II$^*$ AALs 
indicate a density of $\sim$300 cm$^{-3}$ in the Si$^+$ gas. 
\ {\bf (2)} If the gas is photoionized, this density 
implies a distance 
of $\sim$28 kpc from the quasar. Several 
arguments suggest that all of the lines form at approximately 
this distance, with a range of densities determining the range of 
ionizations. 
\ {\bf (3)} Strong Mg I AALs identify neutral gas 
with very low ionization parameter and high density. 
We estimate $n_H > 5\times 10^4$ cm$^{-3}$ 
in this region, compared to only $\sim$15 cm$^{-3}$ where the N V lines form. 
\ {\bf (4)} The total column density is 
$N_{\rm H} < 4\times 10^{18}$ cm$^{-2}$ in the 
neutral gas and $N_{\rm H}\sim 2\times 10^{20}$ cm$^{-2}$ in the moderately 
ionized regions (up to Al III, Si IV, etc.). There may be larger 
column densities in more highly ionized gas, however, the total column 
of $N_{\rm H}\sim 2\times 10^{20}$ cm$^{-2}$ is consistent with 3C 191's  
strong soft X-ray flux and the implied absence of 
soft X-ray absorption.
\ {\bf (5)} The total mass in the AAL outflow is 
$M\sim 2\times 10^9$ M$_{\odot}$, 
assuming a global covering factor (as viewed from the quasar) 
of $\sim$10\% .
\ {\bf (6)} The absorbing gas only partially covers 
the background light 
source(s) along our line(s) of sight, requiring absorption in small 
clouds or filaments $<$0.01 pc across. The ratio $N_H/n_H$ 
implies that the clouds have radial (line-of-sight) thicknesses $<$0.2 pc. 
\ {\bf (7)} The characteristic flow time of the absorbing gas 
away from the quasar is $\sim$$3\times 10^7$ yr.

\section{Discussion}

The physical connection between the quasar 3C 191 and its AALs is 
established by the excited-state lines (Si II$^*$). However, the 
absorber--quasar distance is very large, $\sim$28 kpc. 
Other quasar AALs are known to form much closer to the central 
engines, possibly within a few pc in outflows similar to the BALs 
(Hamann et al. 1997, Barlow \& Sargent 1997, Barlow, Hamann 
\& Sargent 1997). 3C 191 might contain a 
different class of absorber (e.g. much farther from the active nucleus) 
than the majority of AGNs discussed at this meeting. In particular, 
3C 191 does not follow the trend identified by Brandt et al (2000) for
small X-ray to UV continuum flux ratios ($\alpha_{ox}$) accompanying 
large C IV absorption equivalent widths. The outflowing material in 
3C 191 might be leftover from a 
blowout associated with a nuclear starburst, the onset of quasar 
activity, or a past broad absorption line (BAL) wind phase. 
The flow time of $\sim$$3\times 10^7$ yr might therefore 
represent the time elapsed since the formation of the quasar and/or 
an accompanying starburst episode. 

We propose a model for the 3C 191 absorber (Figure 4) in which 
pockets of dense neutral gas (represented by Mg I $\lambda$2853) 
are surrounded by a diffuse, 
spatially extended medium of generally higher ionization 
(e.g. C IV and NV). 
The diffuse clouds contain most of the total column density; 
their greater size and/or greater numbers 
lead to more complete coverage in both velocity and 
projected area. More work is needed to characterize fully this 
class of distant AAL absorbers (see also Hamann \& Brandt 2001, 
Barlow et al. 1997, Tripp et al. 1996, Morris et al. 1986), and 
understand its relationship (if any) to other absorption phenomena 
in quasars/AGNs. 
\smallskip

We are grateful to the staff of the Keck Observatory for their willing 
assistance. FH thanks Bassem Sabra for useful discussions and acknowledges 
support from NSF grant AST 99-84040. CBF acknowldges NSF grant 
AST 98-03072.

\begin{figure}
\plotone{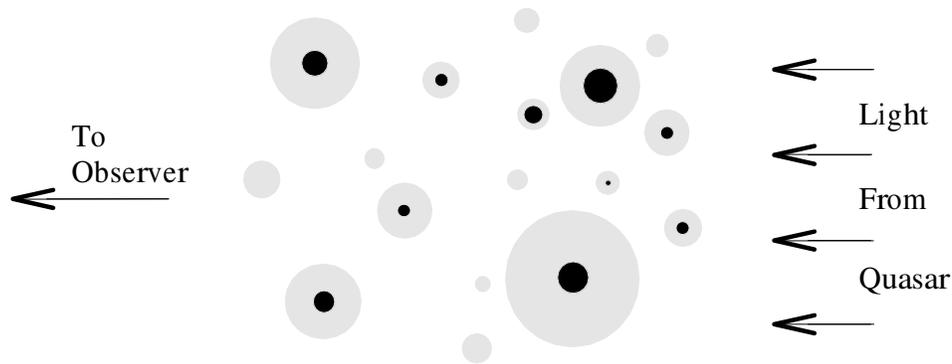}
\caption{Schematic representation of the AAL environment, 
showing pockets of dense neutral gas (filled black circles) 
surrounded by a less dense and more highly ionized medium (grey 
circles). The more extended regions lead to smoother AAL 
profiles and more complete line-of-sight coverage of the background 
light source.}
\end{figure}

\end{document}